\documentclass[prl,amsmath,amssymb,superscriptaddress,twocolumn]{revtex4}

\usepackage{amsmath}    % need for subequations
\usepackage{amsfonts}
\usepackage{amssymb}
\usepackage{graphicx}   % for figures
\usepackage{dcolumn}
\usepackage{bm}

\begin{document}

\title{Exact relativistic kinetic theory of an electron beam-plasma system:\\
hierarchy of the competing modes in the system parameter space}

\author{A. Bret}\email{antoineclaude.bret@uclm.es}
 \affiliation{ETSI Industriales, Universidad de Castilla-La Mancha, 13071 Ciudad Real, Spain}
 \affiliation{Instituto de Investigaciones Energ\'{e}ticas y Aplicaciones Industriales, Campus Universitario de Ciudad Real,
 13071 Ciudad Real, Spain.}
\author{L. Gremillet} \email{laurent.gremillet@cea.fr}
\author{D. B\'enisti}\author{E. Lefebvre}
\affiliation{D\'epartement de Physique Th\'eorique et Appliqu\'ee,
CEA/DIF, BP 12, 91680
Bruy\`eres-le-Ch\^atel, France}

%\date{\today }

\begin{abstract}
Besides being one of the most fundamental basic issues of plasma physics, the stability analysis of an electron beam-plasma
system is of critical relevance in many areas of physics. Surprisingly, decades of extensive investigation had not yet
resulted in a realistic unified picture of the multidimensional unstable spectrum within a fully relativistic and kinetic
framework. All attempts made so far in this direction were indeed restricted to simplistic distribution functions and/or
did not aim at a complete mapping of the beam-plasma parameter space. The present paper comprehensively tackles this problem
by implementing an exact linear model. We show that three kinds of modes compete in the linear phase, which can be classified
according to the direction of their wavenumber with respect to the beam. We then determine their respective domain of preponderance
in a three-dimensional parameter space. All these results are supported by multidimensional particle-in-cell simulations.
\end{abstract}

\maketitle

Relativistic electron beams moving through a collisionless background plasma are ubiquitous in a variety of physical
systems pertaining, e.g., to inertial confinement fusion \citep{Tabak}, the solar corona \citep{Solar}, electronic
pulsar winds \citep{Pulsar}, accreting black hole systems \citep{Microquasar}, active galactic nuclei \citep{AGN}
or $\gamma$-ray burst sources \citep{Piran2004}. Regardless of its setting, the electron beam-plasma system has been
known to be prone to collective processes since the very early studies on the so-called two-stream instability
\citep{BhomGross}, which is of electrostatic nature and therefore characterized by wave and electric field vectors
both parallel to the beam flow direction. Later on, another class of unstable modes was discovered: usually referred
to as ``filamentation'' \citep{Fried1959} or Weibel modes \citep{Weibel}, they are mostly electromagnetic, purely
growing and develop preferentially in the plane normal to the beam. Finally, unstable modes propagating obliquely to
the beam were investigated \citep{Bludman} and found to rule the system in case of cold and diluted relativistic
electron beams \citep{fainberg,califano3,BretPoPHierarchie}. To date, the few kinetic approaches investigating the whole
two-dimensional (2-D) unstable spectrum have been restricted to non-relativistic energy spreads \cite{BretPRL2005}
or to waterbag-like distribution functions of questionable validity at high energy spreads \cite{GremilletPoP2007}.
Among the salient features of the oblique modes revealed by the latter studies are their mostly electrostatic
character and usually efficient interaction with both the beam and plasma components. Yet, because of various
limitations, all the models considered so far have failed to provide a complete picture of the hierarchy of the
competing unstable modes as a function of the system parameters. One should stress that addressing this long-standing
issue, which is the main goal of the present work, is a critical prerequisite to understanding the nonlinear aspects
of the beam-plasma interaction. Among these are the instability-induced collective stopping and scattering, whose level
is expected to depend closely on the nature (\emph{i.e.}, electrostatic or electromagnetic) of the underlying processes
\cite{DavidsonPOF1972,ThodePOF18-1552-1975,OkadaJPP-23-423-1980,HondaPRL85-2128-2000,MedvedevAPJ-618-L75-2005,CalifanoPoP9-451-2002,
MendoncaPRL94-245002-2005,AdamPRL97-205006-2006,OkadaPoP14-072702-2007}.

\begin{figure}[tbp]
\begin{center}
\includegraphics[width=0.4\textwidth]{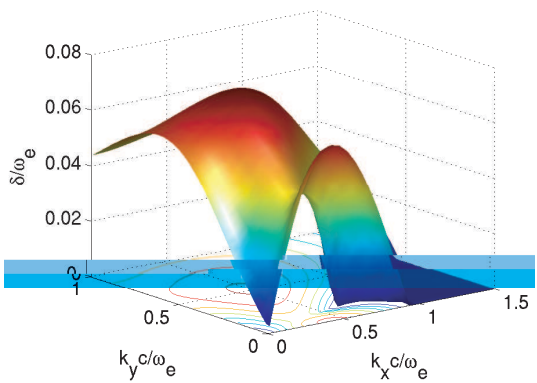}
\end{center}
\caption{(color online). Normalized growth rate in the $(k_x,k_y)$ plane for $n_b/n_p=1$, $\gamma_b=1.2$, $T_b=500$ keV and $T_p=5$ keV. Isocontours are linearly spaced from  0.01 to 0.07.}
\label{fig:1}
\end{figure}

In contrast to past studies, the fully relativistic kinetic model implemented here involves, for both the counterstreaming
beam and plasma populations (hereafter identified by the subscripts $b$ and $p$), unperturbed distribution functions in
the form of drifting Maxwell-J\"{u}ttner functions \cite{Juttner,Wright1975,cubero2007}
\begin{equation}\label{eq:distri}
	f_{\alpha}^{0}(\mathbf{p}) = \frac{\mu_{\alpha}}{4\pi \gamma_{\alpha}^{2}
    K_{2}(\mu_{\alpha}/\gamma_{\alpha})} \exp\big[-\mu_{\alpha}(\gamma(\mathbf{p})-\beta_{\alpha}p_y)\big] \,,
\end{equation}
which allow for arbitrary energy spreads and drifts. Here $\alpha=(b,p)$ stands for the beam or plasma component,
$\beta_{\alpha} = \langle p_{y}/\gamma\rangle$ is the normalized $y$-aligned mean drift velocity, $\gamma_\alpha$ the
corresponding relativistic factor and $\mu_{\alpha}= mc^2/k_B T_\alpha$ the normalized inverse temperature of each
electron component. All momenta are normalized by $mc$. $K_{2}$ denotes a modified Bessel function of the second kind.
Current neutralization is assumed, that is, $n_b \beta_b + n_p \beta_p =0$, where $n_b$ and $n_p$ are the beam and
plasma mean densities, respectively. From now on, the ions form a fixed neutralizing background and collisions are
neglected. It is noteworthy that the above model distribution function is provided a thermodynamically consistent
derivation from first principles in Ref. \cite{Juttner,Wright1975}, where it is shown to maximize the specific entropy
for fixed values of each species' total momentum and energy. In addition to its thermodynamic grounds, this distribution
function is helpful in gaining insight into the stability properties of \emph{smooth} relativistic distributions, as
opposed to the commonly used waterbag distributions \cite{YoonPRA35-2718-1987,SilvaPoP2002,BretPRE2004}, severely
flawed by the neglect of Landau damping. The evolution of the initially homogeneous and unmagnetized system is governed by the relativistic Vlasov-Maxwell
set of equations. Following a routine procedure, the general dispersion equation  for any orientation of the wave vector is derived \cite{Ichimaru,BretPRE2004} and solved numerically. Details of the numerical procedure will be presented elsewhere. Introducing the total electron plasma frequency
$\omega_e$, and normalizing the complex frequency $\omega$ and wave vector $\mathbf{k}$ by $\omega_e$ and $\omega_e/c$ respectively,
there remain four independent variables: the beam mean relativistic factor $\gamma_b$, the beam and plasma temperatures $T_b$
and $T_p$, and the density ratio $n_b/n_p$. A typical map of the growth rate $\delta = \Im \omega $ is displayed in Fig. \ref{fig:1}
for $n_b/n_p=1$, $\gamma_b=1.2$, $T_b$ = 500 keV and $T_p$ = 5 keV. The three aforementioned instability classes are clearly
visible for this configuration. For wave vectors aligned with the beam, the two-stream instability peaks for $k_y \sim 0.5$
whereas, in the perpendicular direction, the fastest growing filamentation mode is found for $k_x \sim 0.5$. Overall, though,
the 2-D unstable spectrum is here dominated by an oblique mode located at $(k_x,k_y) \sim (0.5,0.5)$.

\begin{figure*}[tbp]
\begin{center}
\includegraphics[width=0.95\textwidth]{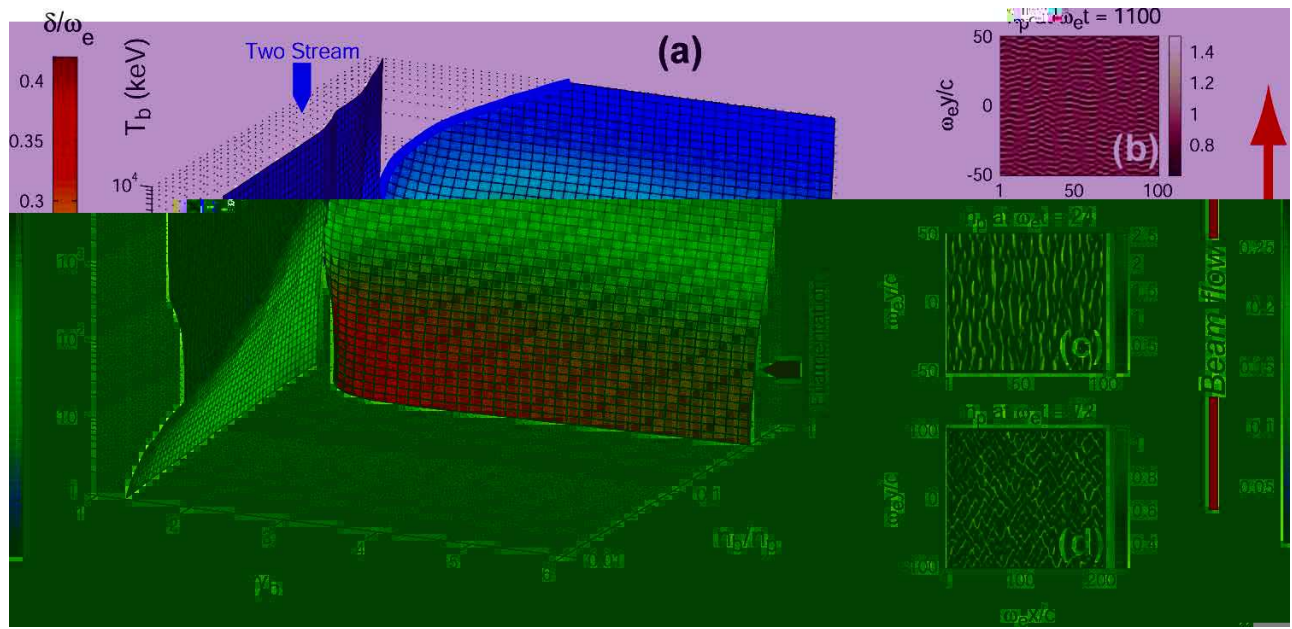}
\end{center}
\caption{(color online). \textbf{Left:} Hierarchy of the unstable modes in the $(n_b/n_p,\gamma_b,T_b)$ parameter space for $T_p$=5 keV. The left surface delimits
the two-stream-dominated domain (at low $\gamma_b$) and the oblique-mode-dominated domain, whereas the right surface delimits the
filamentation-dominated domain (at high $n_b/n_p$) and the oblique-mode-dominated domain.\\
\textbf{Right:} Plasma density profiles at the end of the linear phase as predicted by 2D PIC simulations run with three different
sets of parameters: $n_b/n_p=0.1$, $\gamma_b=1.5$ and $T_b=500$ keV (b); $n_b/n_p=1$, $\gamma_b=1.5$ and $T_b=100$ keV (c);
$n_b/n_p=1$, $\gamma_b=1.5$ and $T_b=2$ MeV (d). In all cases, $T_p=5$ keV. In agreement with linear theory, the three resulting
patterns evidence regimes dominated by two-stream, filamentation and oblique modes, respectively.} \label{fig:2}
\end{figure*}

The three instability classes do not share the same sensitivity to the beam temperature, drift and density. This yields a
non-trivial dependence of the growth rates on the system parameters, hence a varying hierarchy between the instability classes.
Assuming from now on a fixed plasma temperature $T_p=5$ keV, it is possible to determine the regions of predominance of each
instability class in the ($n_b/n_p$,$\gamma_b$,$T_b$) space. The surfaces that delimit regions governed by different instability
classes are displayed in Fig. \ref{fig:2}(a) and colored according to the local maximum growth rate. Points located between the
plane $\gamma_b = 1$ and the left surface define systems dominated by the two-stream instability, while those located between
the right surface and the plane $n_b/n_p =1$ pertain to filamentation-ruled systems. Oblique modes prove to govern the rest of
the parameter space.

The two-stream instability is seen to prevail in the whole non-relativistic range of the beam drift energy ($\gamma_b -1 \ll 1$),
as well as in weakly relativistic systems with hot enough beams. The reason for the latter feature is as
follows: first, filamentation is strongly weakened for weakly energetic beams because its growth rate is proportional to the
beam velocity. Being thus left with the two-stream and oblique modes, we note that the former are handicapped by the relativistic
increase in the longitudinal inertia, while the latter are more sensitive to the beam temperature. As a consequence of these
effects, a system characterized by $n_b/n_p=0.1$, $\gamma_b=1.5$ and $T_b=500$ keV turns out to be mostly subject to the two-stream
instability, as indicated in Fig. \ref{fig:2}(a). We have checked this prediction by means of a 2D particle-in cell (PIC) simulation
using the massively parallel code CALDER \cite{LefebvreNF43-629-2003}. As expected, the plasma density profile, displayed in Fig.
\ref{fig:2}(b) at the time when the instability starts to saturate, exhibits modulations preferentially along the drift direction.
By repeating the Fourier analysis performed in Ref. \cite{GremilletPoP2007}, we have verified that the simulated $\mathbf{k}$-resolved
growth rates closely agree with linear theory.

Let us now consider the filamentation-to-oblique transition, which is found to take place for dense and relativistic beams. The
shape of the corresponding frontier is here more involved than the previous one as it stems from a balance between the three
system parameters. The filamentation growth rate increases with the beam density and decreases with temperature more rapidly than
the oblique mode growth rate. For a cold system \cite{BretPoPHierarchie}, this growth rate scales like $\beta_b/\sqrt{\gamma_b}$, which reaches a maximum for $\gamma_b=\sqrt{3}$. Still in the cold limit, the filamentation-to-oblique frontier reaches a minimum $n_b/n_p\sim 0.53$ for $\gamma_b\sim 3$. The
resulting boundary proves to be mostly determined by $\gamma_b$ for dense, cold and weakly relativistic beams. In the relativistic
and ultra-relativistic regimes, the main parameter is the beam density, although high enough beam temperatures always end up
favoring oblique modes. This result goes against the conventional belief that relativistic systems with $n_b/n_p=1$ are governed
by filamentation \cite{WiersmaAA428-365-2004, CalifanoPRL96-105008-2006}. We show here that this behaviour holds true only as long
as the beam is not too hot. This is illustrated in Fig. \ref{fig:2}(c,d) which compares the simulated plasma density profiles for
two values of the beam temperature, respectively below and above the filamentation-to-oblique boundary. The other parameters are
$n_b/n_p=1$ and $\gamma_b=1.5$. In accord with linear theory, filamentation modes are seen to prevail when $T_b=100$ keV, whereas
oblique modes clearly take over when $T_b=2$ MeV, yielding a characteristic knitted-like pattern.

\begin{figure}[tbp]
\begin{center}
\includegraphics[width=0.45\textwidth]{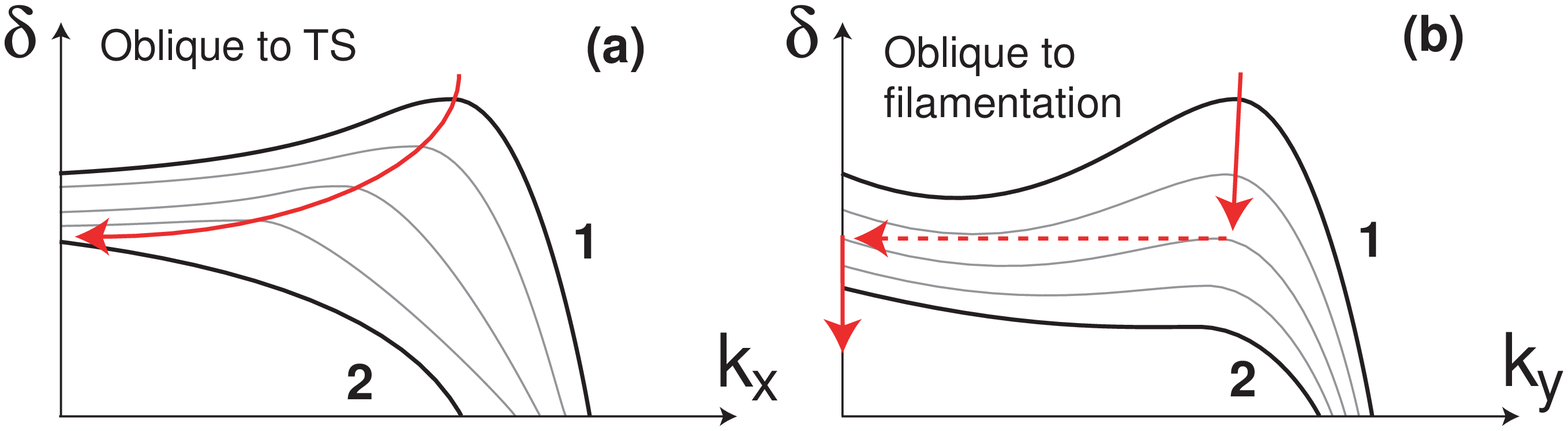}
\end{center}
\caption{(color online). Typical evolution of the fastest growing mode (red arrow) for the oblique/two-stream (a) and oblique/filamentation (b) transitions, of which only the former is continuous.}
\label{fig:3}
\end{figure}

When crossing one of the transition surfaces displayed in Fig. \ref{fig:2}(a), a phase velocity discontinuity may occur as regards the
dominant mode. During the oblique/two-stream transition, the $k_y$ component of the most unstable mode remains almost constant while the $k_x$ component steadily decreases down to zero [Fig. \ref{fig:3}(a), curve 2]. The oblique/two-stream transition is therefore continuous, notably as regards the resulting phase velocity change. During the oblique/filamentation transition [Fig. \ref{fig:3}(b)], by contrast, the $k_x$ component hardly varies, whereas the $k_y$ component evolves as follows. Let us start from a system ruled by an oblique mode with $k_y \neq 0$ and $ \Re \omega\neq 0$ [Fig. \ref{fig:3}(b), curve 1]. As the system moves to the filamentation regime, the $k_y$ component of the dominant mode remains almost constant while its growth rate gradually weakens. When the latter eventually becomes smaller than that of the main filamentation mode (with $k_y = \Re \omega = 0$), the system experiences a sudden change regarding its dominant wave phase velocity, which then switches from a finite value to zero.

\begin{figure*}[tbp]
\begin{center}
\includegraphics[width=0.95\textwidth]{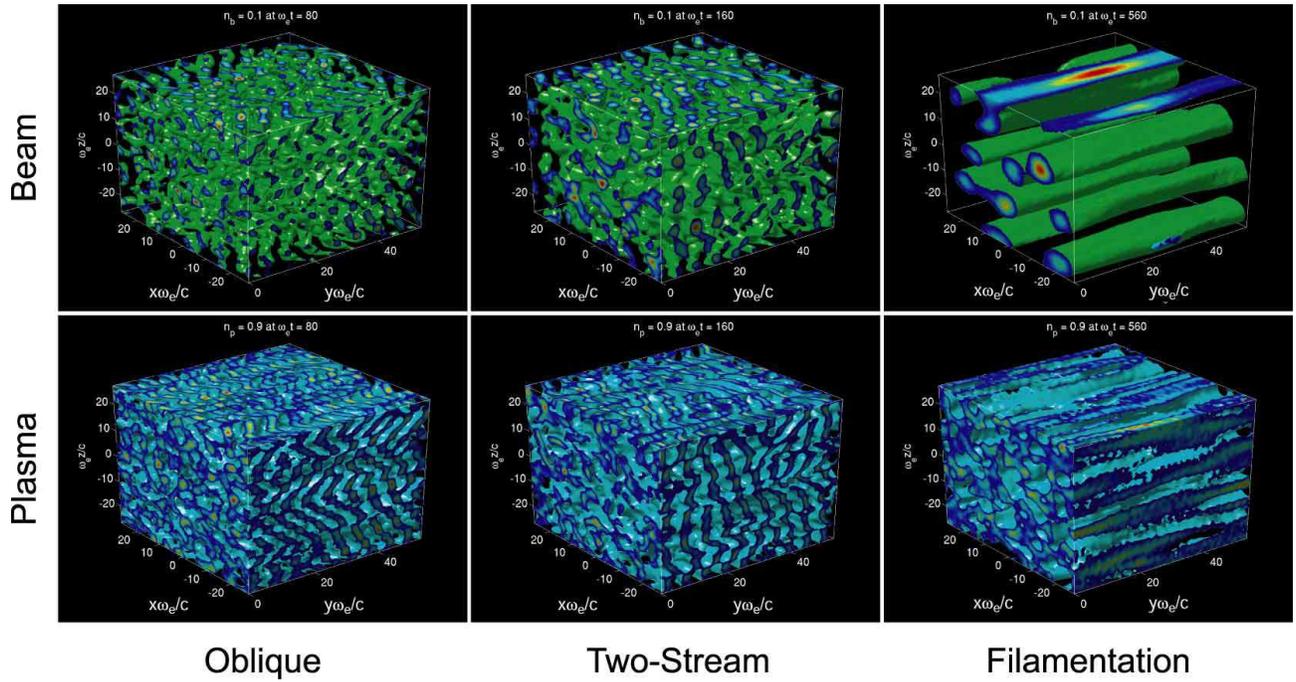}
\end{center}
\caption{(color online). 3D PIC simulation initialized with $n_b/n_p=0.1$, $\gamma_b=3$, $T_b=50$ keV and $T_p=5$ keV: isosurfaces of the beam and plasma
density profiles at $\omega_et=80$, 160 and 560. The system runs through three successive phases, each governed by a distinct instability
class.} \label{fig:6}
\end{figure*}

So as to gain insight into the field generation within a more realistic geometry, we have performed a three-dimensional PIC simulation
of a diluted beam-plasma system defined by $n_b/n_p=0.1$, $\gamma_b=3$, $T_b=50$ keV and $T_p=5$ keV (Fig. \ref{fig:6}). In agreement
with Fig. \ref{fig:2}(a), the system is initially governed by oblique modes, as confirmed by the beam and plasma profiles at $\omega_et=80$.
Later on ($\omega_{e}t=160$), though, the system gets ruled by two-stream modes.  A rough, quasilinear-like argument supporting this
transition can be given assuming that the beam distribution function has retained a Maxwell-J\"{u}ttner form.  At $\omega_{e}t=160$ the
best fit is obtained for $\gamma_b \sim 1.6$ and $T_b\sim 200$ keV. Because of the low beam density, we can reasonably neglect the changes
in the plasma distribution function. Figure \ref{fig:2}(a) then indicates that the system is indeed dominated by two-stream modes. After
saturation of all electrostatic modes, the system eventually reaches a filamentation regime, owing to a remaining, high enough anisotropy,
as already observed in Ref. \cite{GremilletPoP2007}. Yet, this transition cannot be explained in light of the present linear model because
the distribution functions then strongly depart from the form (\ref{eq:distri}).

Let us conclude by making two important remarks.  Firstly, given our model distribution function (\ref{eq:distri}), complete linear
stabilization of the system can never be achieved in the whole parameter space. In particular, the maximum filamentation growth rate
can be shown to scale like $1/T_b^{3/2}$ in the large $T_b$ limit. Secondly, the oblique/filamentation boundary in the plane $T_b = 1$ keV
behaves like $n_b/n_p \sim 1-0.86\gamma_b^{-1/3}$ when $\gamma_b \gg 1$. As a result, the filamentation-ruled region is all
the more squeezed against the plane $n_b/n_p=1$ when the beam energy is high. This also implies that the ultra-relativistic regime is
asymptotically oblique unless $n_b/n_p=1$.  This could bear important consequences in a number of astrophysical scenarios which involve
relativistic factors up to $10^3$ and beyond \cite{Dieckmann,Aharonian,Medvedev1999}.

Thanks are due to Claude Deutsch and Marie-Christine Firpo for enriching discussions. This work has been partially achieved under projects FIS 2006-05389 of the Spanish Ministerio de Educaci\'{o}n y Ciencia, and PAI08-0182-3162 of the Consejer\'{i}a de Educaci\'{o}n y Ciencia de la Junta de Comunidades de Castilla-La Mancha. The simulation work was performed using the computer facilities of CEA/CCRT.

\end{document}